\documentclass{book}
\usepackage{krantz_single}%
\usepackage{amsfonts}
\usepackage{graphicx}

\input epsf		%  files

\title{Quantum Phase Transitions in quasi-one dimensional systems}
\author{Thierry Giamarchi}

\begin{document}
\maketitle
\frontmatter

\pagestyle{myheadings} \markboth{QPT in Q1D systems}{Table of
Contents}

\tableofcontents

\mainmatter \markboth{QPT in Q1D systems}{QPT in Q1D systems}

\vspace*{-1cm}

\chapter{Quantum Phase Transitions in Quasi-one-dimensional Systems}
\label{chapter:giamarchi}

Among the various systems, one dimensional (1D) and quasi-one dimensional
(quasi-1D) systems are a fantastic playground for quantum phase
transitions (QPTs), with rather unique properties. There are various
reasons for that special behavior.

First, purely 1D systems are rather unique.
Contrary to their higher-dimensional counterparts~\cite{TG:mahan_book}, interactions play a major role since in 1D particles cannot avoid the effects of interactions.
This transforms any individual motion of the particles into a
collective one. In addition to these very strong interaction
effects, in 1D the quantum and thermal fluctuations
are pushed to a maximum, and prevent the breaking of continuous
symmetries, making simple mean-field physics inapplicable. The
combination of these two effects leads to a very special
universality class for interacting quantum systems, known as
Luttinger liquids (LLs)~\cite{TG:haldane_bosonisation_spin}.

I will not review here all the aspects of LL
physics since many such reviews exist, but refer the reader in
particular to~\cite{TG:giamarchi_book_1d} for a complete
description of this area of physics, with the same notations as the
ones used in the present chapter. For what concerns us, the
important point is that we take the LL to be in a critical phase,
in which correlations decrease, at zero temperature, as power
laws of space and time. This makes the system extremely fragile
to external perturbations and leads to a host of QPTs. Examples of such perturbations are the effects of
a lattice, which leads to a Mott transition, and disorder that
leads to localized phases such as Anderson localization or the
Bose glass. Each of these transitions is characterized by a
quantum critical point (QCP) that can be computed from  LL theory.
The 1D nature of LLs has other consequences: the excitations
can fractionalize.  In particular, an excitation such as
adding an electron can split into several collective
excitations, such as one carrying spin but no charge, called a \textit{spinon}
and one carrying charge but no spin, called a \textit{holon}.  I will not dwell
on this physics of purely 1D systems that is now
well characterized and  refer the reader to~\cite{TG:giamarchi_book_1d} for this aspect of QCPs in purely 1D systems.

Although  purely 1D physics and QPTs are by now rather well under control from the
theoretical point of view, there is a category of perturbations
that is still at the frontier of our theoretical knowledge.
These are the perturbations that are produced by the coupling
of several 1D systems.  Then, when
one parameter, for example the temperature or the inter-chain coupling, is varied, the system crosses over from a 1D situation with  exotic
LL physics, to the more conventional high dimensional one.  How one can reconcile such different
physical limits, for example recombining the spinons and holons to
re-form an electron, to perform such a dimensional crossover is
a very challenging and still open question. Such questions are
not only important on the theoretical side but have direct
applications to experimental systems such as organic~\cite{TG:chemicalreview} or inorganic~\cite{TG:wang_purple_arpes}
superconductors, spin chains and ladders~\cite{TG:tennant_kcuf_1d,TG:lake00_kcuf3,TG:dagotto_ladder_review}
and cold atomic systems~\cite{TG:bloch_cold_lattice_review} (see also Chap.~\ref{chapter:bloch}) which provide
realizations of such coupled 1D systems.

Quasi-1D systems thus leads to their own
interesting sets of QCPs, and these
are the ones on which I will focus in the present chapter. I
will start by examining the simple case of coupled spin chains
and ladders, then move to the case of bosons, and finally deal
with the more complicated and still largely open case of
fermions.

\section{Spins: From Luttinger Liquids to Bose-Einstein Condensates}

The simplest example of coupled 1D system is
provided by coupled spin systems (see e.g.
\cite{TG:tennant_kcuf_1d,TG:lake00_kcuf3} for experimental systems).
In addition to their own intrinsic interest and their direct
experimental realization, they will also serve to illustrate
several important concepts that will be directly transposed
with increasing complexity to the case of bosons and fermions.

Coupling chains starting from 1D is a highly nontrivial process. Going from one spin chain to two, called the \textit{spin ladder problem}, already leads to non-trivial
physics. Indeed, although spin-1/2 systems are gapless the
coupling of two spin-1/2 chains leads to the formation of a
spin gap,  similar to the Haldane gap that occurs for
integer spins~\cite{TG:haldane_gap}. I will not
discuss this physics in details since it is by now well
established and covered in several textbooks and refer the
reader to the literature on the subject
\cite{TG:dagotto_ladder_review,TG:giamarchi_book_1d}.

Here, I consider the case when an infinite number of low
dimensional units are coupled. As can be readily understood the
physics will depend crucially on the fact that the systems that
got coupled are already in a critical state (such as spin-1/2
chains) or whether they have a gap (such as spin dimers, spin 1
chains or two legs ladders). These two cases are the prototypes
of QCP in coupled 1D systems and we will examine
them separately

\subsection{Coupled Spin-1/2 Chains}
\label{TG:sec:coupledSpinOneHalfChains}

An isolated spin-1/2 chain is described by a LL. As can be
expected in 1D, no long range order can exist. However, the
spin-spin correlation functions decay as a power law, at zero
temperature, indicating the presence of quasi-long range order.
Focusing on the case of the antiferromagnetic exchange, which
is the natural realization in condensed matter systems, spin-spin correlations decay as~\cite{TG:giamarchi_book_1d}
\begin{equation} \label{TG:eq:LLpower}
 \langle S^+(x)S^-(0)\rangle \propto (-1)^x \left(\frac1x\right)^{1/(2K)}, \quad
 \langle S^z(x)S^z(0)\rangle \propto (-1)^x \left(\frac1x\right)^{2K},
\end{equation}
where $K$ is the LL parameter and depends only on the spin
exchange anisotropy between the $XY$ and $Z$ plane,
$J_Z/J_{XY}$. For an isotropic Heisenberg interaction $K =1/2$,
 both correlations decrease as $1/r$, up to logarithmic
corrections. Temperature cuts this power-law decrease and
transforms it into an exponential decay of the correlation
beyond a scale of order $u \beta$, where $\beta$ is the inverse
temperature and $u$ the velocity of spin excitations.

The inter-chain coupling introduces a term of the form
\begin{equation} \label{TG:eq:interspin}
 H_\perp = J_\perp \sum_{\langle\mu\nu\rangle} \int dx S_\mu(x) \cdot S_\nu(x),
\end{equation}
where $\langle \mu,\nu \rangle$ denotes two neighboring chains
$\mu$ and $\nu$. Because the spin is an object that admits a
good classical limit, one can analyze the physics of such a
term in a mean-field approximation by assuming that the spin on
each chain acquires an average value, for example in the $Z$
direction. This allows one to decouple (\ref{TG:eq:interspin}) and
transform it to an effective Hamiltonian corresponding to a
self-consistent staggered magnetic field applied on a single
chain $H_\perp \simeq J_\perp \sum_{\nu} h_{\rm eff} \int dx
(-1)^x S_{\nu,Z}(x)$. Using the standard bosonization
representation of the spins, the Hamiltonian then becomes a
sine-Gordon Hamiltonian whose sine term represents the effects
of the effective staggered field~\cite{TG:schulz_coupled_spinchains}. The
physics of such a Hamiltonian is well known, and there are two
phases. First, there is a critical phase, where one recovers the massless
excitations. This corresponds to the high temperature phase
where the chains are essentially decoupled. Second, there is a massive phase
where the cosine is relevant\footnote{In the standard language of renormalization theory, terms are defined as \textit{relevant} when they do not tend to zero under a renormalization transformation, and \textit{irrelevant} otherwise.} and acquires an average value.
This means that $\langle (-1)^x S_Z(x) \rangle$ is now non-zero, which signals true long range order in the system.  The system
thus exhibits a genuine phase transition as a function of the
temperature towards an ordered state that would correspond to
 anisotropic antiferromagnetic three dimensional (3D) behavior.
The critical temperature can be  analyzed by using
scaling analysis of the inter-chain coupling. Using
(\ref{TG:eq:LLpower}) leads to the renormalization flow of the
inter-chain exchange:
\begin{eqnarray} \label{TG:eq:rgspin}
 \frac{d J_{XY}}{dl} & = & J_{XY}(2- \frac1{2K}), \\
 \frac{d J_{Z}}{dl} & = & J_{Z}(2- 2K),
\end{eqnarray}
where $l$ describes the renormalization of the bandwidth of the
systems $\Lambda(l) = \Lambda_0 e^{-l}$ and $\Lambda_0$ is the
bare bandwidth. One sees that one of the couplings is always
relevant regardless of the value of $K$ and that one has always
an ordered state (for a non frustrated inter-ladder coupling) at
low enough temperature.

\begin{figure}[t]
\vspace*{0.5cm}
 \centerline{\includegraphics[width=0.7\textwidth]{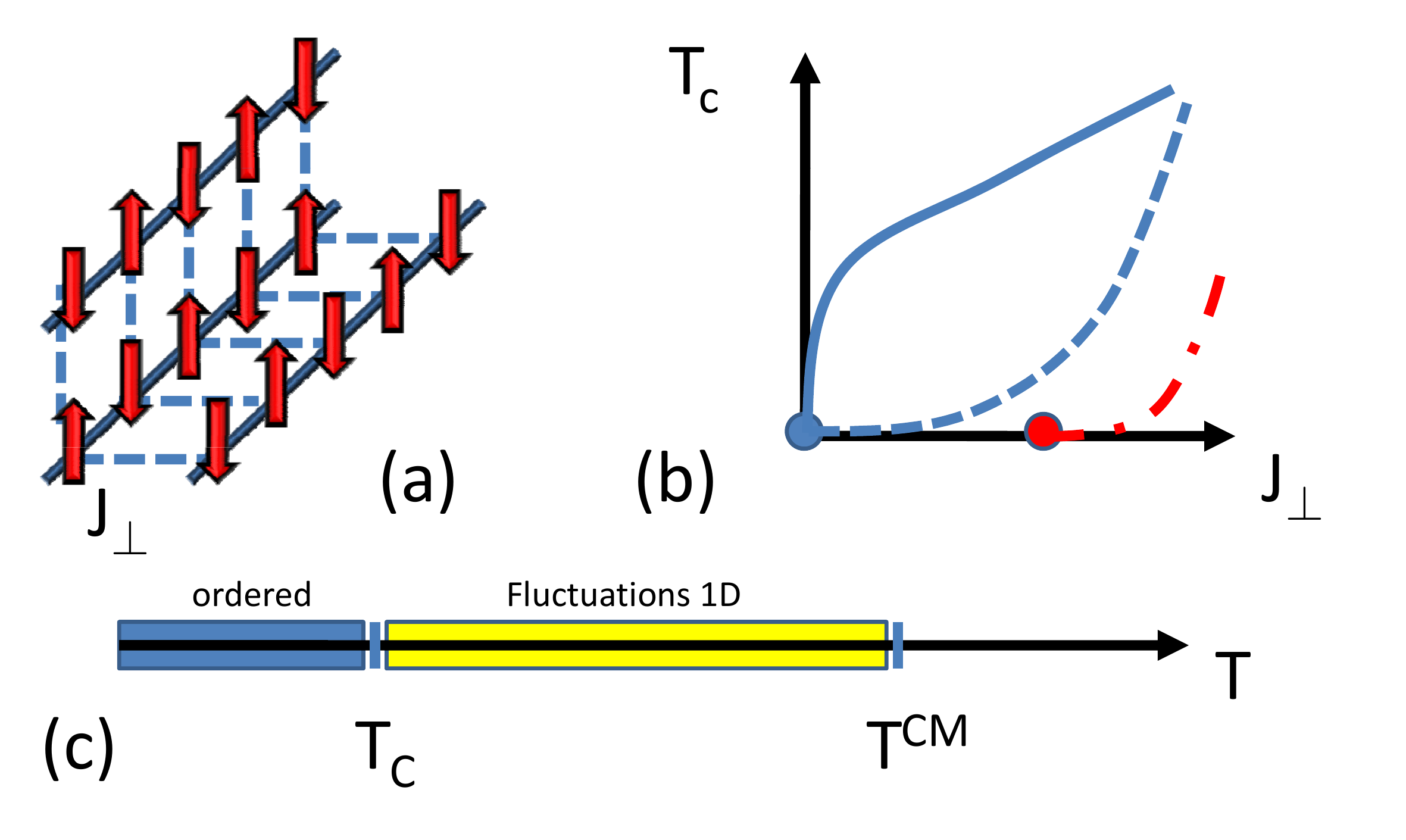}}
 \caption{(a) Coupled 1D chains. The inter-chain coupling $J_\perp$ (dashed line)
 is much weaker than the intra-chain  one $J_\parallel$ (full line). This leads to properties linked to
 the Luttinger liquid ones of the 1D chains. (b) Depending on the Luttinger liquid parameter
 $K$, the critical temperature is a power law of the inter-chain coupling $J_\perp$, since the
 coupling is strongly renormalized by 1D fluctuations. The exponent is either smaller than one (full line), or when the fluctuations increase, larger
 than one (dashed line). If $K$ is below (or above depending on the correlations, see text) a certain value,
 fluctuations are small enough and lead to an ordered state as soon as some inter-chain coupling is introduced.
 However, if the fluctuations are large enough, ordering is suppressed unless the inter-chain coupling
 reaches a critical value (dashed-dotted line). Note that in this case usually another correlation orders since
 several instabilities are in competition. (c) Because of the strong 1D fluctuations, the mean-field
 temperature can be quite different from the actual critical temperature to an ordered state.}
 \label{TG:fig:1dfluct}
\end{figure}

However, the critical nature of the 1D
systems leads to a strong renormalization of the critical
temperature with respect to a naive mean-field approximation;
the latter would result in $T_c \sim J_\perp$. Instead, strong 1D fluctuations lead to
\begin{equation}
T_c^{XY} = J_\parallel \left(\frac{J_\perp}{J_\parallel}\right)^{1/(2-1/2K)}, \quad\quad
T_c^{Z}  =  J_\parallel \left(\frac{J_\perp}{J_\parallel}\right)^{1/(2-2K)},
\end{equation}
as can be deduced directly from (\ref{TG:eq:rgspin}), since the
critical temperature follows from $l^* =
\log(\Lambda_0/T_c)$ for which the running coupling is $J_\perp(l^*) \sim
J_\parallel$. The strong 1D fluctuations thus
have a large effect on the critical temperature and create a wide
regime where the system is dominated by 1D fluctuations, as
indicated on Fig.~\ref{TG:fig:1dfluct}.
Although the temperature scale is strongly affected, the
critical behavior still corresponds to that of the
higher-dimensional case. However, the quasi-1D nature of
the problem has strong consequences for the existence
of extra modes of excitations in comparison to what happens for
a more isotropic system~\cite{TG:schulz_coupled_spinchains}. We will
come back to this point when discussing  bosons, where these
modes can be more simply understood.

\subsection{Dimer or Ladder Coupling}

A much more complex behavior occurs when the objects that become
coupled have a gap in their spectrum, a gap that is in
competition with the presence of the inter-chain coupling (\ref{TG:eq:interspin}). In this case one can expect a real
QPT to occur in which the system goes
from a low-dimensional gapped situation, to a higher-dimensional
ungapped one. This transition is called generically a \textit{deconfinement transition}, since the system
changes both its effective dimensionality and the nature of its
spectrum at the same time.  This particular type of QCP manifests itself in several types of systems and
we will examine it for  spins, bosons and fermions.

\begin{figure}
 \centerline{\includegraphics[width=0.7\textwidth]{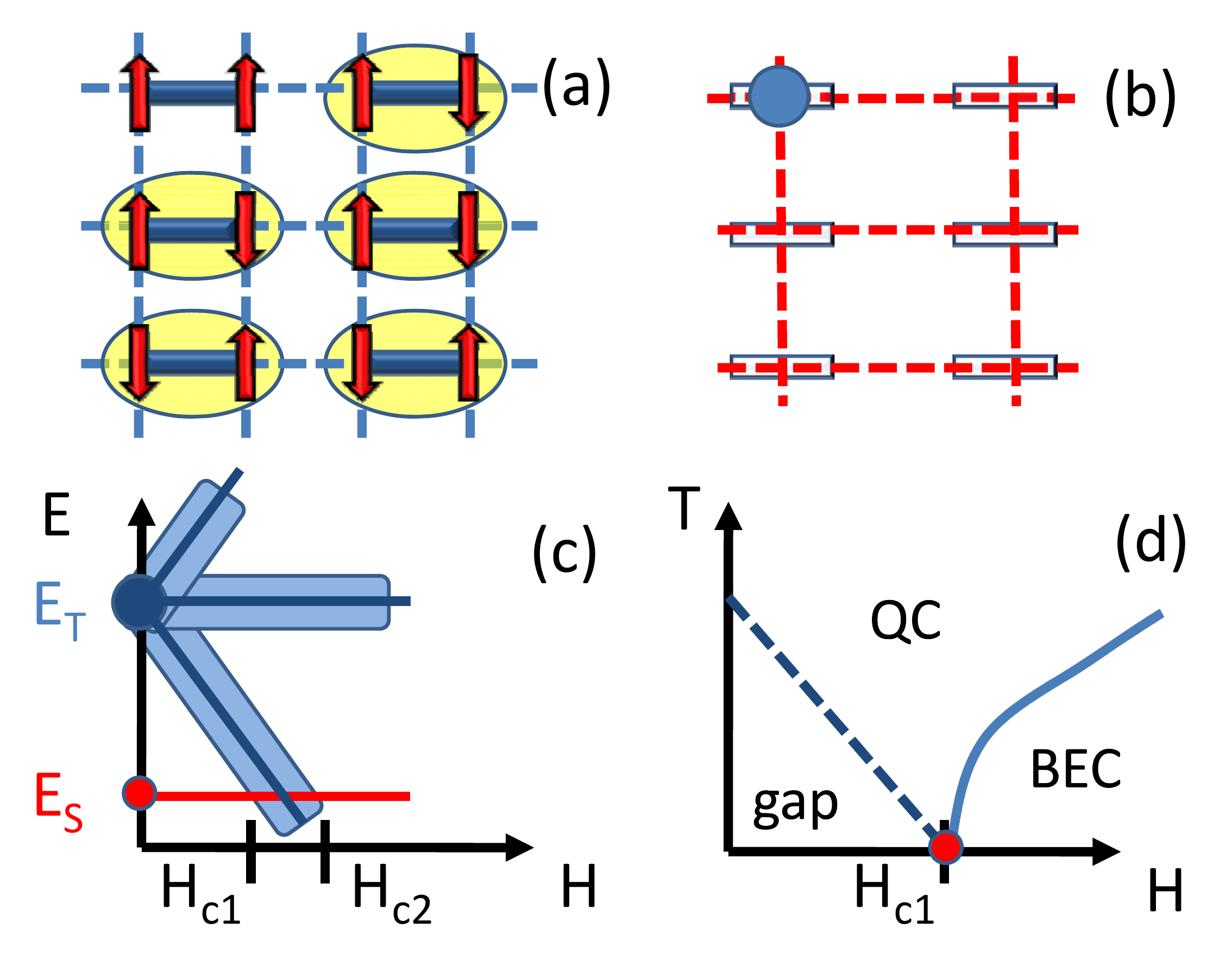}}
 \caption{(a) If one of the exchanges $J_d$ (thick line) is larger than the others $J$ (dashed line)
 then one has a system made of coupled dimers. Because a dimer goes into a singlet state (oval shape), the ground
 state of such system is made of decoupled singlets. (b) One can map such a system onto a system of hard-core bosons, the presence
 of a boson denoting a triplet state on the dimer, and its absence a singlet. Because of  spin exchange one has an equivalent
 system of bosons hopping on a lattice with a kinetic energy given by the inter-dimer magnetic exchange $J$. In addition to the hard-core constraints bosons have  nearest-neighbor interactions. (c) Application of a magnetic field lowers the energy
 of one of the triplet states (thick dot) compared to the singlet one (small dot). Because the triplet disperses, there is a band
 of triplet excitations (triplons). Changing the magnetic field allows one to fill this band of triplons which are the hard-core bosons
 of (b). The field $H_{c1}$ corresponds to the first triplon entering the system, while the field $H_{c2}$ is a filled band of triplons.
 Such a system thus provides an excellent venue in which to study interacting bosons on a lattice, since the density of bosons can be controlled
 directly by the magnetic field, and measured by  the magnetization along the field direction. (d) this system has a quantum
 phase transition at $H_{c1}$; a similar transition exists at $H_{c2}$, not shown here. The triplons exhibit  Bose-Einstein condensation (BEC), which corresponds
 in  spin language to antiferromagnetic order in the direction perpendicular to the magnetic field. ``Gap'' and ``QC'' denote
 the gapped state in which there are no triplons and the quantum critical state, respectively.}
 \label{TG:fig:triplet}
\end{figure}

The case of spin is the simplest.  To illustrate the nature
of this QCP let us consider first the case of a system made of
dimers, weakly coupled by (\ref{TG:eq:interspin}). In this case
each dimer has a gap between a singlet state and the three
triplet states. The gap is of order $J_d$, the dimer spin
exchange. Since there is a gap in the spin excitation spectrum,
the dimer is robust to the inter-dimer exchange coupling and the
ground state in the case $J_\perp \ll J_d$ is made of
essentially uncoupled dimers. In this case we are considering a
cluster of zero-dimensional objects coupled by the inter-dimer
coupling. If we now place the system in a magnetic field the
dimer gap reduces and ultimately the lowest triplet state reaches the level of the singlet one as depicted in
Fig.~\ref{TG:fig:triplet}.
In this case the inter-dimer coupling is able to delocalize the
triplets and lead to a transition where the system will go from
a set of essentially uncoupled zero-dimensional objects to an
essentially 3D antiferromagnet. Quite remarkably,
this deconfinement transition can be analyzed by mapping the
singlet-lowest triplet onto a hard-core boson. The system is
thus equivalent to a set of hard-core bosons, the density of which is
controlled by the magnetic field.\footnote{Zero boson density means that
all dimers are in the singlet state, while one boson per site
means that each dimer is fully polarized.} When the first
triplets enter the system at a critical field $h_{c1}$ the
bosons are extremely dilute;  thus their hard-core
interaction is not felt very strongly. The phase transition is
thus  Bose-Einstein condensation~\cite{TG:giamarchi_coupled_ladders}. The QCP at
$T=0$ corresponds to the point where the chemical potential is
such that a finite density of bosons starts to appear. This has
several interesting consequences for the nature of the phase
diagram and in particular allows one to predict features such as
the critical temperature which behaves as
\begin{equation}
 T_c \propto (h - h_{c1})^{2/d},
\end{equation}
as well as non-monotonous temperature dependence of the
magnetization of the system. Since its original prediction, these
behaviors have been studied and observed in several compounds
with both 3D~\cite{TG:nikuni_bec_tlcucl,TG:ruegg_bec_tlcucl,TG:matsumoto04_tlcucl3}
and bi-dimensional structure
\cite{TG:sebastian_BEC_dimensional_reduction}. For a review on
these aspects I refer the reader to
\cite{TG:giamarchi_BEC_dimers_review}.

\begin{figure}[t]
 \centerline{\includegraphics[width=0.7\textwidth]{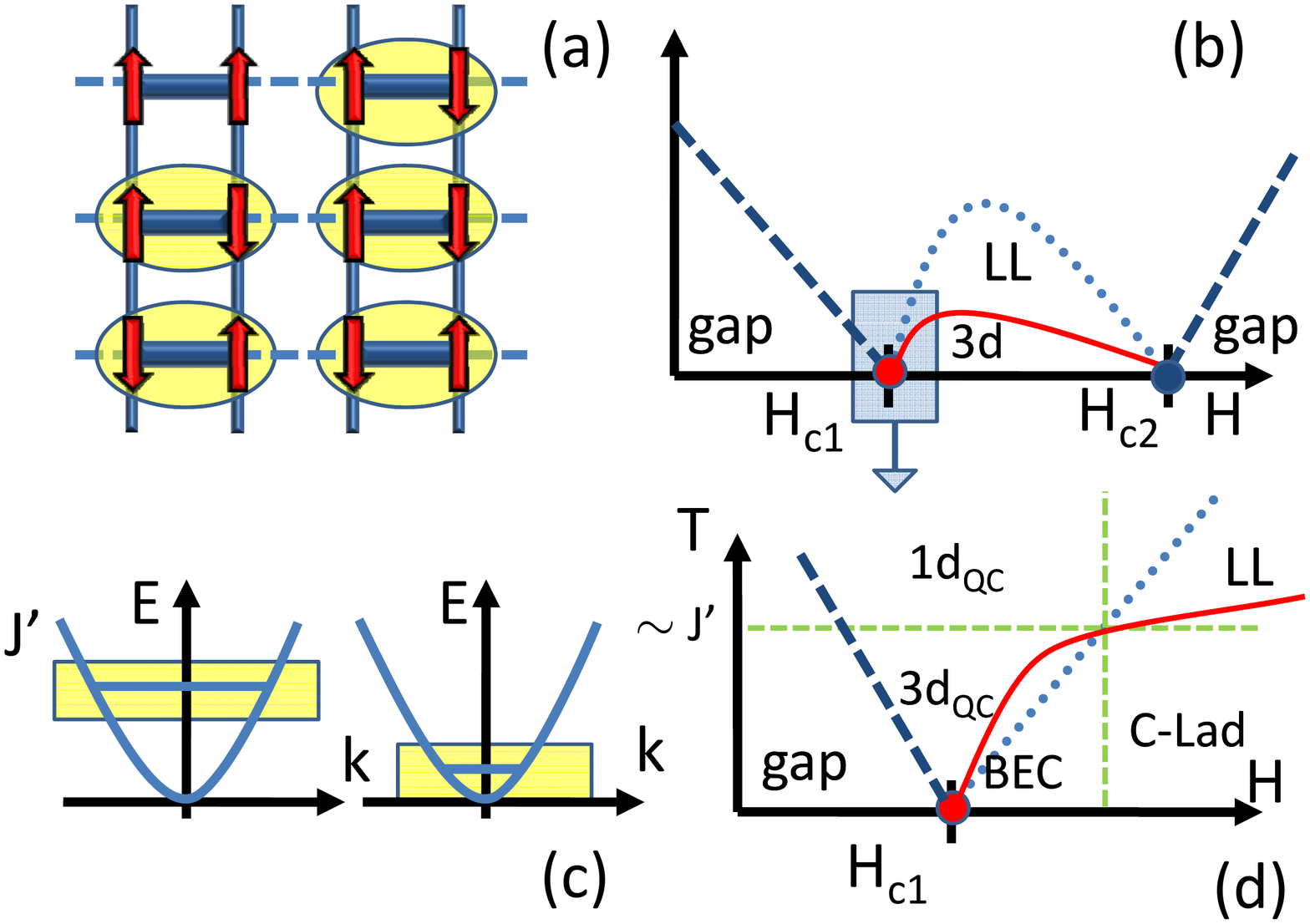}}
 \caption{(a) Coupled ladders correspond to a hierarchy of coupling where the leg coupling $J_\parallel$
 (solid line) is smaller than the rung coupling $J_d$ (thick line), but larger than the inter-ladder coupling
 $J'$ (dashed line). In the same way as in Fig.~\ref{TG:fig:1dfluct} this leaves room for 1D fluctuations
 and Luttinger liquid physics to modify the behavior as compared to the more isotropic case of Fig.~\ref{TG:fig:triplet}. (b) As a result,
 a LL regime exists between $H_{c1}$ and $H_{c2}$. In this regime a good physical description of the triplons is to consider
 that they behave  as spinless fermions. The regime $3d$ where the spins order antiferromagnetically is the equivalent
 of the BEC regime of Fig.~\ref{TG:fig:triplet}. (c) As long as the coupling $J'$ is larger than the chemical potential $H-H_{c1}$ one can
 consider the system as made of coupled 1D systems and 1D fluctuations play a major role. On the contrary, when
 $J' > (H-H_{c1})$ one must consider a 3D system from the start since there is no room for 1D fluctuations to take place.
 One thus returns to the case of Fig.~\ref{TG:fig:triplet}. (d) As a result there is a more complex crossover regime when the field
 $H$ gets close to $H_{c1}$. C-Lad denotes coupled ladders and BEC is the Bose-Einstein condensation of Fig.~\ref{TG:fig:triplet}. }
 \label{TG:fig:1d3dcross}
\end{figure}

A similar class of deconfinement transitions occurs when the
objects that are coupled have a 1D structure. For
example, when one deals with a spin one chain
\cite{TG:affleck_field} or a two leg spin ladder
\cite{TG:giamarchi_coupled_ladders}. Both these structures are
characterized by a gap. Application of a magnetic field allows one to break the gap and to study the transition to the 3D behavior. In this case one gets a very interesting
behavior which is depicted in Fig.~\ref{TG:fig:1d3dcross}.
If one is far from the critical field $H_{c1}$ the chemical
potential of the excitations in the 1D chain is
high and in particular larger than the inter-chain coupling. One
is thus dealing essentially with the situation depicted in the
previous section, of weakly coupled 1D LL, and one
can study the transition to the ordered state. Dimers present
several advantages to study this phase transition since the
singlet is extremely robust to external perturbations, such as
dipolar interactions. Such interactions would break the spin
rotation symmetry in the $XY$ plane and thus in the boson
mapping break the phase $U(1)$ symmetry. Recently, very nice
experimental realizations of ladder systems have been analyzed
\cite{TG:watson_bpcb,TG:klanjsek_nmr_ladder_luttinger,TG:ruegg_specific_heat_ladder,TG:thielemann_neutron_ladder}.
These analyses have allowed for a quantitative test of the predictions of the LL and
of the generic scenarios described above for the transition to
the ordered state.

When getting closer to the field $H_{c1}$ one cannot consider
that the system is made of coupled LLs since the temperature is
getting larger than the distance to the bottom of the band, and
one has to consider the 1D quantum critical behavior. Such a
situation, although more complex, can still be analyzed by
various techniques~\cite{TG:orignac_nmr_bec}. In a similar way,
when lowering the temperature one has to consider an additional
crossover where the temperature becomes  smaller than the
inter-ladder coupling $J_\perp$. This corresponds to going from
a 1D quantum critical regime of weakly coupled
ladders to the 3D one of coupled dimers, as
described in Sec.~\ref{TG:sec:coupledSpinOneHalfChains}. The resulting physical
behavior is thus quite complex and largely not
understood, despite the analysis of
several quantities, such as the NMR response.
Other ways to control such a phase transition include applying
pressure instead of the magnetic field; this
changes the ratio $J_\perp/J_\parallel$ and thus makes the
system more 3D, as described in Chap.~\ref{chapter:sachdev}.  We proceed to consider
this type of transition in more detail for the related but
different case of itinerant bosons and fermions.

\section{Bosons: From Mott Insulators to Superfluids}
\label{TG:sec:coupledSpinChains}

Consider first the case
of coupled bosonic chains.  In principle, such a system is  very
close to the problem of coupled spin chains, since a spin 1/2
can be represented by a hard-core boson. However, the absence of
the hard-core constraint, and of the nearest neighbor
interaction that corresponds to the $J_Z \sum_i S^Z_i
S^Z_{i+1}$, lead in practice to quite different regimes than for
 coupled spin chains. Nevertheless, most of the techniques
and concepts that we used for  spin chains will be directly
useful for  coupled bosonic chains.

Although in principle one can realize coupled bosonic systems
in condensed matter, e.g. using Josephson junction arrays~\cite{TG:fazio_josephson_junctions_review}, it is relatively
difficult to obtain a good realization. Recently, cold atomic
systems in optical lattices have provided a remarkable and very
controlled realization on which many of the aspects discussed
below can be  tested in experiments, as discussed in Chap.~\ref{chapter:bloch}.

For the case of bosons the coupling between the chains comes
mostly from the single-particle hopping from one chain $\mu$ to
the neighboring one $\nu$.  This term in the Hamiltonian takes the form
\begin{equation}
 H_\perp = - t_\perp \sum_{\langle\nu,\mu\rangle} \int dx \, \psi^\dagger_\mu(x) \psi_\nu(x).
\end{equation}
Along the chains, interacting bosons can be represented either
by a continuum theory or directly on a lattice by a
Bose-Hubbard model. In 1D both these cases can be
mapped to a LL description for the low energy properties~\cite{TG:giamarchi_book_1d,TG:cazalilla_1d_bec,TG:giamarchi_bosons_salerno}.
The corresponding Hamiltonian is
\begin{equation} \label{TG:eq:llbosons}
 H^0 = \frac1{2\pi} \int dx \left[\frac{u}{K} \pi\Pi(x) + u K (\nabla\theta)^2\right],
\end{equation}
where $\theta$ is the superfluid phase determined by $\psi(x) =
\rho(x)^{1/2}e^{i\theta(x)}$. The field $\Pi(x)$ is canonically
conjugate to $\theta$ and is associated with density fluctuations.
The long-wavelength density fluctuations can
indeed be represented as $\rho(x) = \rho_0 -
\pi^{-1}\nabla\phi(x)$ and $\pi\Pi(x) = \nabla\phi(x)$. Also, $u$ is
the velocity of the sound waves in the 1D systems, while $K$ is
the LL exponent which depends on the microscopic interactions.
For bosons with a contact interaction, $K=\infty$ for
non-interacting bosons, while $K\to 1^+$ when the contact
interaction becomes infinite.  The latter is the \textit{Tonks-Girardeau limit}.

As for spins, the properties will be crucially dependent on
whether the 1D bosonic system is gapless or not. In the
continuum, the full description of the system is indeed given
by (\ref{TG:eq:llbosons}). In the presence of a lattice or a periodic
potential one should take into account oscillations of the
density with periodicity $2\pi\rho_0$, of the form
$\delta\rho(x) \propto \cos(2\pi\rho_0 x - 2\phi(x))$. These
oscillations, when commensurate with the period of the lattice, e.g. with one boson per site, can lead to a Mott-insulating
phase for the bosons~\cite{TG:haldane_bosons,TG:fisher_boson_loc,TG:giamarchi_book_1d,TG:giamarchi_bosons_salerno}.
In this case the full 1D Hamiltonian becomes
\begin{equation} \label{TG:eq:mott}
 H = H^0 - g \int dx \cos(2\phi(x)),
\end{equation}
where $g$ is a constant proportional to the lattice strength
for small lattices or the interaction for large ones
\cite{TG:giamarchi_book_1d}. Such a term becomes relevant for $K <
2$ and leads to an ordered phase $\phi(x)$. Since $\phi$ is
locked by the cosine term, this corresponds to frozen density
fluctuations, and thus a Mott-insulating phase with an integer
number of bosons per site. Since $\phi$ and $\theta$ are
conjugate variables this implies that  superfluid
correlations decrease exponentially and that the quasi-long-range superfluid order is destroyed. Let us examine both these
cases, with and without the commensurate term.

\subsection{Coupled Superfluid: Dimensional Crossover}
\label{TG:sec:coupledSuperfluid}

the case in the absence of the lattice, or when the lattice is
irrelevant, is very similar to the coupled spin chains examined
in Sec.~\ref{TG:sec:coupledSpinChains}.  Each 1D chain is critical with a
quasi-long-range superfluid order, since with
(\ref{TG:eq:llbosons}) the superfluid correlations decay as a
power-law:
\begin{equation}
 \langle \psi(x) \psi^\dagger(0) \rangle \propto x^{-1/2K}.
\end{equation}
One can treat the inter-chain coupling in the mean-field approximation, since boson single-particle operators can have a mean-field value:
\begin{equation}
 H_\perp = - t_\perp \sum_{\langle\mu\nu\rangle} \int dx \left[\langle \psi^\dagger_\mu(x)\rangle \psi_\nu(x) + h.c.\right] \to - \Delta \int dx \cos(\theta(x)),
\end{equation}
where $\Delta = 2 z \rho_0^{1/2} t_\perp \langle
\psi^\dagger(x) \rangle$, and $z$ is the coordination of the
lattice. Thus one finds a sine-Gordon Hamiltonian in the
superfluid phase that can freeze the phase $\theta$ and lead to
 {\it long range} superfluid order~\cite{TG:efetov_coupled_bosons,TG:ho_deconfinement_short}.
Note that without the mean-field approximation the interaction
term in  phase language becomes
\begin{equation} \label{TG:eq:josephson}
 H_\perp = - 2 t_\perp \rho_0 \sum_{\langle\nu,\mu\rangle} \int dx \cos(\theta_\nu(x) - \theta_\mu(x)).
\end{equation}
Given the quadratic form of (\ref{TG:eq:llbosons}), and taking
time as an extra classical dimension, one can immediately map
this problem onto coupled $XY$ planes. For a 3D system one would thus be in the universality class of the five-dimensional $XY$ model,
 justifying the use of the mean-field approximation.\footnote{The mean-field approximation is exact in the limit that the number of nearest neighbors approaches infinity.}  As for the problem of spins,  1D
fluctuations strongly renormalize the critical temperature
as compared to the naive mean-field value $T_c \sim t_\perp$. The scaling is very similar to
that of Sec.~\ref{TG:sec:coupledSpinChains}~\cite{TG:ho_deconfinement_short}.

An interesting effect can be seen when looking at
fluctuations around the ground state in the low temperature
superfluid phase. As can be readily seen by performing a random-phase-approximation (RPA)
treatment~\cite{TG:ho_deconfinement_short} of
the Hamiltonian $H^{1D} + H_\perp$, two eigenmodes exist. One is
the standard phase mode, where the amplitude of the order
parameter is essentially fixed but $\theta_\nu(x)$ slowly
varies in space and from chain to chain. The energy of this
mode goes to zero, since this is the standard Goldstone mode of
the superfluid. However, another eigenmode exists,
corresponding to a change in amplitude of the order parameter, and thus also $\rho(x) = \rho_0 + \delta\rho(x)$. Not
surprisingly, this mode is dispersing above a finite energy
$E_0$ but exists as a sharply defined mode, in a way very
similar to plasmons in charged systems. Such a mode would
not appear in a more isotropic superfluid, as can be readily
seen by solving the Gross-Pitaevskii equation~\cite{TG:pitaevskii_becbook}. This is one clear-cut case
where the higher-dimensional system still shows some traces of
its 1D origin and displays qualitative differences as compared to an
isotropic system. Such modes have also been observed close to Mott
transitions in isotropic systems~\cite{TG:huber_amplitude_mode_cold,TG:ho_deconfinement_short}.

\subsection{Coupled Mott Chains: Deconfinement Transition}
\label{TG:sec:coupledMottChains}

As for  spins, the situation is much more interesting and
complex when the 1D chains are in the Mott-insulating phase. In
this case it is clear that there is a competition between the
Mott term (\ref{TG:eq:mott}), caused by the periodic potential
along the chains that prefers order in the phase $\phi$
controlling the density, and the inter-chain Josephson term
(\ref{TG:eq:josephson}) that prefers to order the superfluid phase
$\theta$. This is the bosonic equivalent of the competition
between the spin gap and the transverse magnetic order that
existed for the spin chains, and was discussed in Sec.~\ref{TG:sec:coupledSuperfluid}.  This competition leads to a deconfinement QPT.

In contrast to the case of  spins, where the 1D gap was closed
by  changing the magnetization, in the case
of bosons  one stays at a
commensurate density and the critical point is reached by
changing the strength of the inter-chain hopping. This is very
similar to the question of the application of pressure in the
case of the spins.  A similar transition to the one studied in Sec.~\ref{TG:sec:coupledSuperfluid} could also occur in the case of
bosons.  It corresponds to the application of a chemical
potential taking the system away from the commensurate point.
In this case the 1D system is described by a
commensurate-incommensurate phase transition~\cite{TG:giamarchi_book_1d} and the universality class of the
deconfinement transition is different. We confine our present discussion to
the commensurate case; see~\cite{TG:ho_deconfinement_short} for
the incommensurate one.

\begin{figure}[t]
 \centerline{\includegraphics[width=0.5\textwidth]{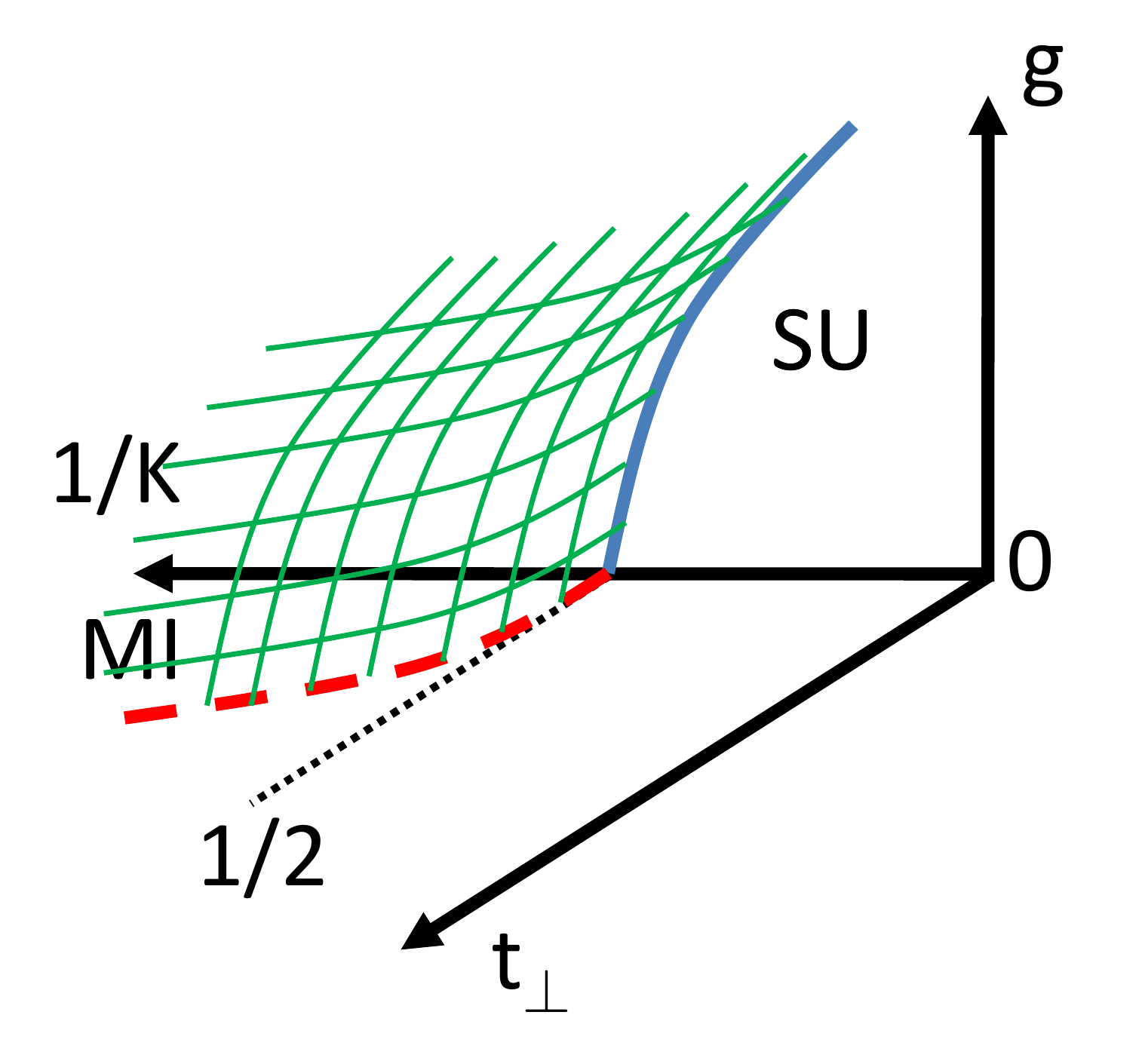}}
 \caption{Phase diagram for quasi-1D bosons on a lattice (at $T=0$): $g$ is the intra-chain
 periodic potential responsible for the Mott transition  for commensurate filling; $t_\perp$ is the inter-chain   kinetic energy or Josephson coupling; and $K$ is the Luttinger parameter that depends on the intra-chain  interactions ($K =\infty$ corresponds to free bosons).
  The thick solid line is the boundary between a 1D Mott insulator and a quasi-ordered 1D superfluid. For very small $g$ the Mott phase occurs for
  $K < 2$. The thick dashed line indicates
  how the extra kinetic energy provided by the inter-chain coupling weakens the Mott state. The green lines are the deconfinement transition
  between a Mott insulator and an anisotropic 3D superfluid.}
 \label{TG:fig:schemphas}
\end{figure}

In the commensurate case there are several ways to analyze the
deconfinement transition. In the mean-field approximation the system is described by a \textit{double sine-Gordon Hamiltonian}. This Hamiltonian has a set of
remarkable properties that have been looked at in various
contexts~\cite{TG:Jose_Kadanoff}.
The critical point can be crudely obtained via
renormalization of the two relevant operators and fixing the phase of the operator that first reaches strong
coupling. A more sophisticated analysis can be found in Ref.~\cite{TG:ho_deconfinement_short}. In
particular, the universality class of the transition can be
shown to be that of the $(d+1)$-dimensional $XY$ model~\cite{TG:fisher_boson_loc}.
Indeed the operator $\cos(2\phi)$ is nothing but the vortex
creation operator for  excitations of the phase $\theta$~\cite{TG:giamarchi_book_1d}. Each chain can thus be mapped onto a
discrete $XY$ Hamiltonian of the form $H = J \sum_{ij}
\cos(\theta_i-\theta_j)$ and the inter-chain coupling has a
similar form but with a different coefficient. A schematic
representation of the phase diagram is shown in
Fig.~\ref{TG:fig:schemphas}.
As for the spins, there is a deconfinement transition between an
essentially 1D insulating phase, where the bosons
are in a Mott state, and an anisotropic 3D superfluid phase. In the Mott phase, there is a gap towards
excitations, and the density is well ordered, i.e. one particle per
site. The anisotropic superfluid phase is gapless and is
 similar to the one that was discussed in the
absence of a lattice in the previous section. Such physics can
be probed in cold atomic systems in systems made of coupled
bosonic tubes such as~\cite{TG:stoferle_tonks_optical}.

\section{Fermions: Dimensional Crossover and Deconfinement}

Let us finally move to the very challenging problem of coupled
1D fermionic chains. As for bosons, the system is
described by a Hamiltonian
\begin{equation}
 H = \sum_\alpha H^{\rm 1D}_\alpha - t_\perp \sum_{\langle\alpha,\beta\rangle} \int dx \, \psi^\dagger_\alpha(x) \psi_\beta(x).
\end{equation}
However, there is a very important difference between the
fermionic case and the two previous sections. Indeed, for
fermions the single-particle operator cannot have an average
value: $\langle \psi_\alpha \rangle = 0$. We cannot
treat the inter-chain coupling by treating the single-particle
operator in a mean-field approximation as we did before. It is
thus difficult to find theoretical tools to tackle this problem
on the analytical side. Similarly, on the numerical side one
cannot use the efficient methods of the 1D world, such as the density matrix renormalization group of Chap.~\ref{chapter:schollwoeck}. One has to use the
arsenal of higher-dimensional Monte Carlo methods, which can suffer from the sign problem, as described in several chapters of Part~\ref{part:4}.

\begin{figure}[t]
 \centerline{\includegraphics[width=0.7\textwidth]{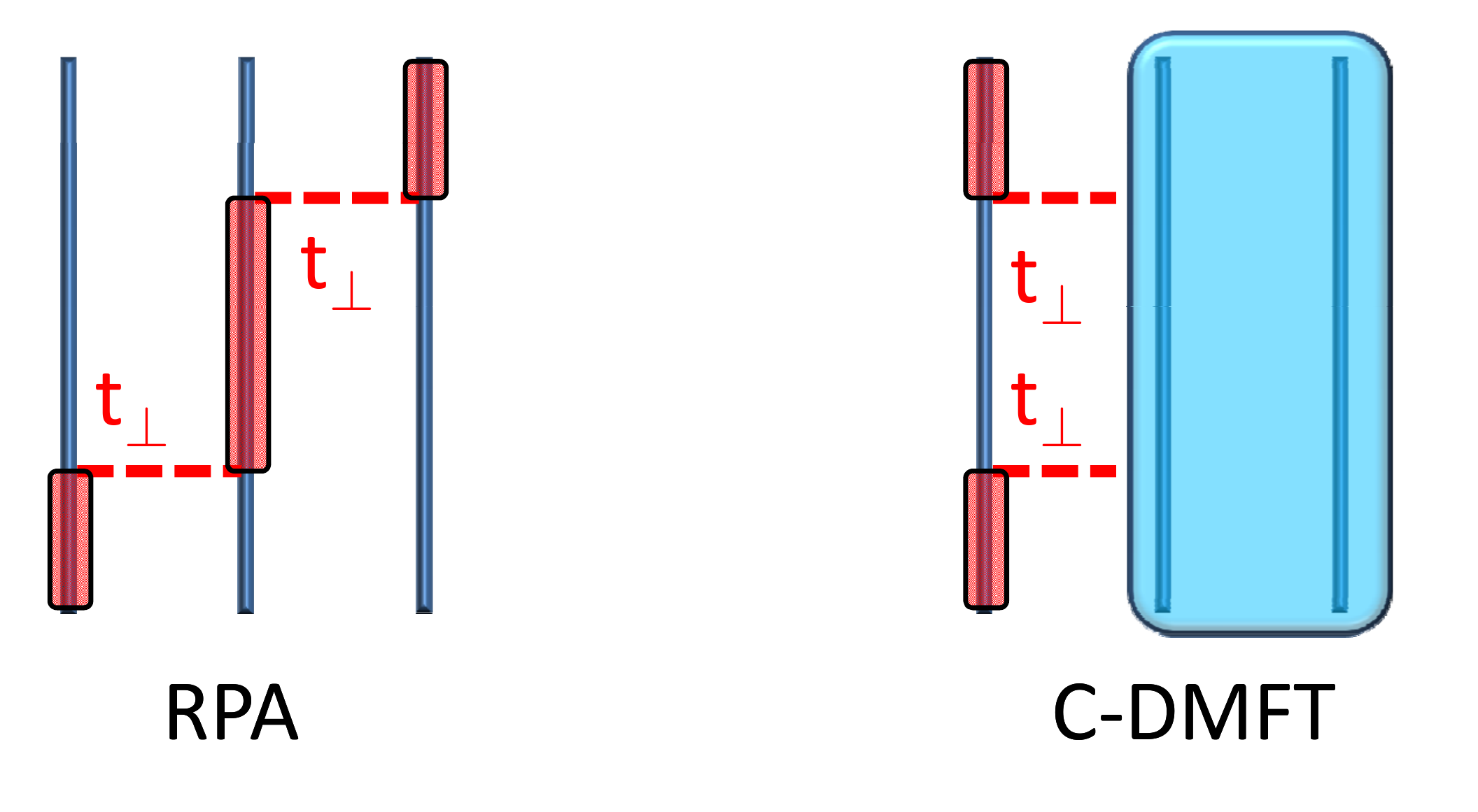}}
 \caption{The two main approximations used to tackle the inter-chain coupling for fermions.
 RPA: the fermions hops but essentially never comes back to the original chain. The properties of a single chain
 are thus not affected at all by the inter-chain hopping. In particular, the Mott gap is strictly independent of $t_\perp$.
 Ch-DMFT: all the chains but one are treated as a self-consistent bath. The 1D Green's function thus depends approximately on
 the inter-chain hopping. This corresponds to the approximation of taking the 1D self-energy independent of the transverse momentum
 $k_\perp$ but potentially dependent on the frequency and momentum along the chains.}
 \label{TG:fig:chdmft}
\end{figure}

Analytically, one can use a renormalization technique similar to
the one introduced in Secs.~\ref{TG:sec:coupledSuperfluid}-\ref{TG:sec:coupledMottChains} to study the
relevance of the inter-chain hopping, as we will discuss in
more detail below. Unfortunately, it will only yield
information about whether the inter-chain coupling is relevant,
not what the strong-coupling fixed point actually is. To understand this
physics, and replace the mean-field treatment used in the two
previous cases, two approximate methods have proven useful for
 fermions, as summarized in Fig.~\ref{TG:fig:chdmft}. The first and
simplest one is to treat the inter-chain coupling in RPA~\cite{TG:essler_rpa_quasi1d}. This leads to
\begin{equation}
 G(k,k_\perp,\omega) = \left(G^{-1}_{\rm 1D}(k,\omega) - t_\perp \cos(k_\perp)\right)^{-1}.
\end{equation}
RPA has the advantage of being very simple.
However, it neglects all feedback of the inter-chain hopping on
the 1D properties themselves, which is clearly a
very brutal approximation. A better approximation is provided
by an extension of  dynamical mean-field theory (DMFT)~\cite{TG:georges_dmft}, where one treats all the chains but one as
an external self-consistent bath into which the particles can
jump~\cite{TG:arrigoni_tperp_resummation,TG:biermann_oned_crossover_review}.
This is detailed in
Fig.~\ref{TG:fig:chdmft}.
>From a more formal point of view one could view it naively as
performing a mean-field approximation on the second-order term
in the hopping in the action
\begin{eqnarray}
 S'_\perp &=& t_\perp^2 \sum_{\mu,\nu} \int dr dr' \psi_\mu^\dagger(r) \psi_\nu(r) \psi_\nu^\dagger(r') \psi_\mu(r') \nonumber \\
 & & \to t_\perp^2 \sum_{\mu,\nu} \int dr dr' \psi_\mu^\dagger(r) \langle \psi_\nu(r) \psi_\nu^\dagger(r')\rangle \psi_\mu(r'),
\end{eqnarray}
where $r=(x,\tau)$ are the space-time coordinates. One thus has an effective intra-chain  kinetic energy,
nonlocal in space and time,  whose amplitude is controlled
by the single-particle Green's function on another chain
$G_\nu(r,r')$. Such a Green's function must thus be determined
self-consistently, and one has to solve an effective 1D Hamiltonian with a modified kinetic energy. Because
the energy now depends self-consistently on the results of the
inter-chain tunneling, there is a direct feedback of the
inter-chain tunneling on the 1D features, contrary to the
case of the RPA. Although still imperfect this is an improvement
which should allow one to obtain several of the features more
accurately.

As for bosons, let us examine the two cases depending on whether
the 1D system is critical (LL) or gapped
(typically a Mott insulator). In the first case the inter-chain hopping leads simply to a dimensional crossover between 1D and higher-dimensional behavior, while in
the second case a deconfinement transition occurs.

\subsection{Dimensional Crossover}

If the 1D system is in a critical LL state, the
inter-chain hopping has in general a strong influence on it. In
a similar way as for the bosons, one can estimate the
relevance of the single-particle hopping by a simple scaling
analysis. If the single-particle Green's function decreases as
\begin{equation}
 G^{\rm 1D}(r,t) \propto r^{-[K_\rho + K^{-1}_\rho + 2]/4}t^{-[K_\rho + K^{-1}_\rho + 2]/4},
\end{equation}
where $K_\rho$ is the charge LL parameter
\cite{TG:giamarchi_book_1d}, then the perpendicular hopping obeys
the renormalization equation coming from the second order
expansion in the inter-chain hopping~\cite{TG:brazovskii_transhop,TG:schulz_moriond}:
\begin{equation}
 \frac{\partial t_\perp}{\partial l} = t_\perp \left[2 - \frac{1}{4}\left(K_\rho + K^{-1}_\rho + 2\right)\right].
\end{equation}
Thus when $K_\rho + K^{-1}_\rho > 6$ the inter-chain hopping is
irrelevant. The intra-chain interactions are enough to prevent
the coherent hopping since a single-particle excitation must be
reconstructed for the electron to be able to hop from one chain
to the next. Note that this implies rather strong, as well as
finite range interactions~\cite{TG:giamarchi_book_1d}. Indeed, for
a purely local interaction such as the one coming from a
Hubbard model, $1/2 < K_\rho < 2$ and thus the inter-chain hopping would always be relevant.  The fact that
the inter-chain hopping is irrelevant does not mean that there
is no coupling at all between the chains; it just means that
single-particle excitations cannot propagate coherently between
them. One must then go to second order in the inter-chain hopping. To second order, the inter-chain hopping generates both
particle-hole coupling, i.e. either density-density or spin-spin,
or particle-particle coupling, i.e. Josephson.  One of
these couplings can become relevant and lead to an ordered
state. The couplings can be treated by
mean-field theory, as explained in Secs. and in more detail in Ref.~\cite{TG:giamarchi_book_1d}.

If $K_\rho + K^{-1}_\rho < 6$, called \textit{moderate
interactions}, the inter-chain hopping is a relevant
perturbation. There will thus exist an energy scale\footnote{
E.g. temperature, frequency determined via probes such as optical
conductivity, energy determined via probes such as STM, etc.  For more
details on how to probe this crossover see
\cite{TG:giamarchi_review_chemrev,TG:bourbonnais_review_book_lebed}.}
below which the system will crossover from  1D to higher-dimensional behavior. Note that this is a simple
crossover and that no phase transition occurs here.  The 1D
nature of the system strongly affects the \emph{scale} at which
the inter-chain hopping acts. This scale is roughly determined
by the condition $t_{\perp}(l^*) = 1$. Thus one finds
the crossover at
\begin{equation}
E_{\rm cross} \propto E_F \left(\frac{t_\perp}{E_F}\right)^{2/(2-2\zeta)},
\end{equation}
where $\zeta = [K_\rho + K^{-1}_\rho + 2]/4$ is the single-particle correlation exponent. As we saw for  spins,
interactions  considerably lower the crossover scale and
reinforce the range of validity of the 1D regime.
the properties of the resulting low-temperature phase
is still a largely open question. In particular, how much this
phase remembers the effects of strong correlations coming from
the high energy 1D physics is  important
to determine. Some elements of response can be obtained via the
various mean-field approximations mentioned above. Quantities
strongly depending on the transverse directions are  very
interesting but also very difficult to compute. This includes
the Hall effect and the transverse conductivity. In particular, the latter can be an indication of the
dimensional crossover transition temperature since the absence
in the 1D regime or the presence in the higher-dimensional
regime of well formed single-particle excitations will lead to
very different temperature dependence.  This is the case e.g.
for organics superconductors~\cite{TG:moser_conductivite_1d} and
for inorganic compounds~\cite{TG:wang_purple_arpes}.

\subsection{Deconfinement Transition}

\begin{figure}
 \centerline{\includegraphics[width=0.7\textwidth]{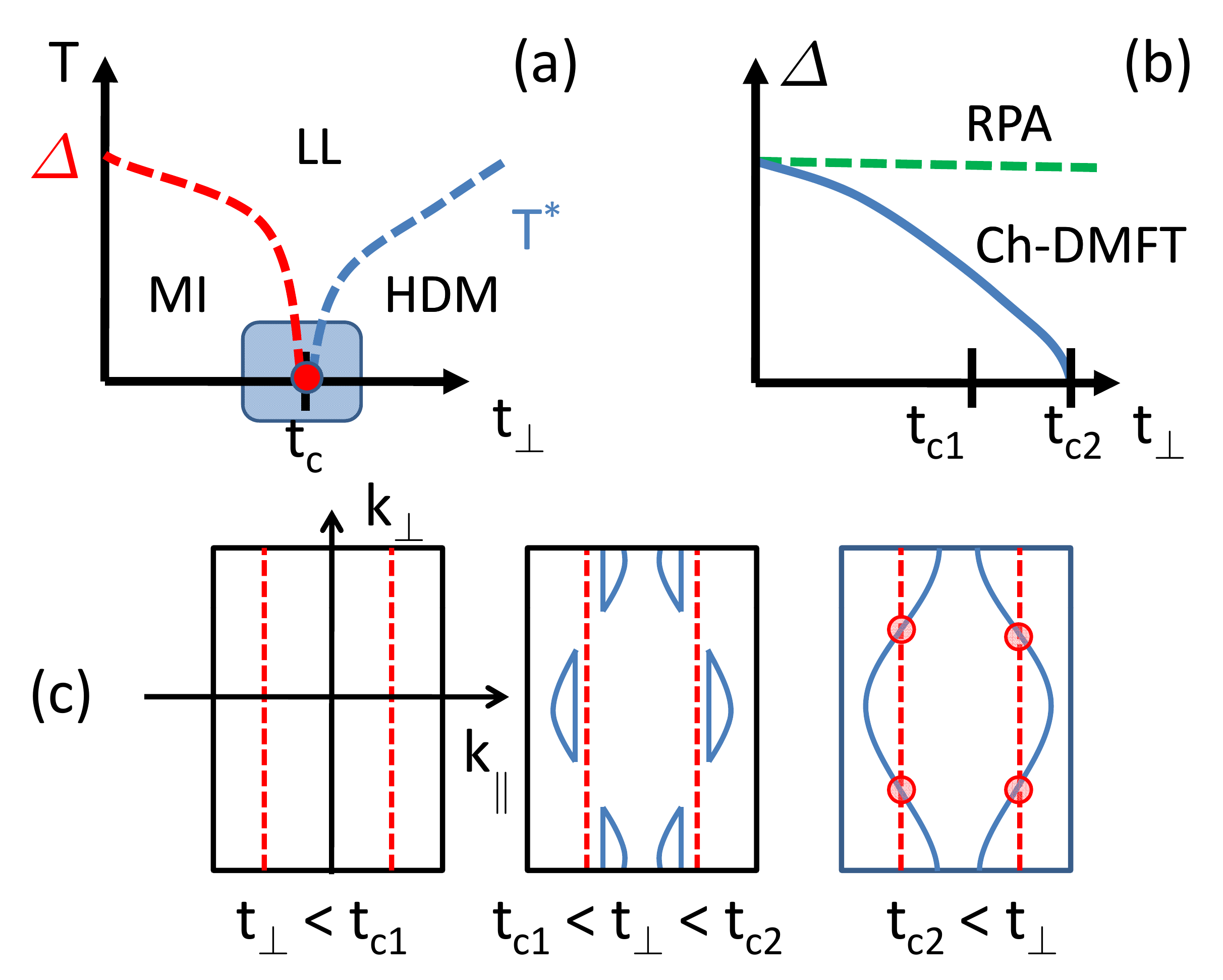}}
 \caption{(a) Phase diagram for the deconfinement transition of quasi-1D coupled fermionic chains. There is a quantum
 critical point (dot) that separates a Mott insulator from a higher-dimensional metal (HDM). For small inter-chain hopping $t_\perp < t_c$
 there is a crossover scale, the renormalized Mott gap $\Delta$. For temperatures $T > \Delta$ the system behaves as a LL, while
 it acts as a Mott insulator for smaller temperatures. On the metallic side there is a coherence scale $T^*$ scaling with the inter-chain hopping
 that separates the LL regime from the higher-dimensional metal in which coherent hopping between the chains occurs. The nature
 of the deconfinement transition, and whether there is a unique transition or a more complex scenario such as two consecutive transitions,
 are still largely open questions. (b) The effective Mott gap $\Delta$ as predicted both by the RPA and the Ch-DMFT approximation. (c) With the current approximations there would be three different phases, leading to different Fermi surfaces. (i) At small gap one has a Mott insulator
 with only zeros of the Green's function.  (ii) At intermediate gaps the inter-chain hopping induces an indirect doping, leading to pockets on the Fermi surface. (iii)
 For larger hopping the gap closes and the pockets join to give back the open Fermi surface of a quasi-1D metal. Note that the Fermi parameter varies
 strongly on such a Fermi surface, reminiscent of the 1D character. In particular, there would be hot spots (dots). The scenario with two transitions can be established
 for spinless fermions with the Ch-DMFT approximation. Whether such a scenario survives with fermions with spins is still an open question. In the
 RPA the gap is not destroyed by the inter-chain hopping, leading to a rigid-band scenario. As a consequence, the pockets depicted
 in (c) never close in such an approximation.}
 \label{TG:fig:deconf}
\end{figure}

The situation is particularly difficult for fermions when the
1D phase is gapped. The expected phase diagram is schematically
indicated in Fig.~\ref{TG:fig:deconf}.
Let us examine in more detail the features of such a
transition. Although the generic shape of the $T-t_\perp$
diagram reminds us of what is to be expected for a generic QPT, the order and even the number of transitions
are not known with certainty. In some cases, both from the RPA and
from Ch-DMFT, two different transitions are  expected to
occur.

The physics of the massive (Mott) phase is relatively clear.
The effective gap is reduced by the additional kinetic energy
provided by the inter-chain hopping. Such an effect is well
described by the Ch-DMFT approximation, which  shows a
reduction of the 1D gap as $t_\perp$ increases. In principle, one
needs the transverse directions to be on a non-bipartite lattice. Otherwise, the Fermi surface remains nested
despite the inter-chain hopping and the gap does not vanish.
This is specially important  for the organic
compounds. At a certain critical value of the inter-chain hopping, one expects to break the Mott-insulating phase and
recover a higher-dimensional metal. How this transition occurs is
still unclear. Within RPA and Ch-DMFT approximations, one goes
through an intermediate phase where pockets appear as depicted
in Fig.~\ref{TG:fig:deconf}. At larger values of the inter-chain hopping
the pockets merge and one expects to recover an open Fermi
surface. This scenario is established for spinless particles
\cite{TG:berthod_dmft_spinless}. Whether it survives for fermions
with spins is still an open question.

Another important open question is to get an accurate
description of the properties of the higher-dimensional metal,
and in particular whether one gets back a Fermi liquid or
whether there is a serious influence of the strong correlations
that existed in the 1D part of the phase diagram.
Even if one recovers a Fermi liquid, as is the case with the
Ch-DMFT method for example, there is clearly a strong
variation of the lifetime and quasiparticle weight along the
Fermi surface (hot spots). How to reliably compute such effects
is a considerable challenge.

Finally, let us point out that such transitions are important
for a host of quasi-1D systems. Deconfinement
transitions are investigated in  organics~\cite{TG:bourbonnais_review_book_lebed,TG:pashkin_deconfinement_tmtsf}
but the inter-chain hopping also clearly plays a crucial role in systems
such as purple bronze~\cite{TG:wang_purple_arpes}.
Both these systems still have a poorly understood superconducting
phase in the higher-dimensional regime. How much such a phase is
influenced  by the 1D nature of the material
still remains to be determined. Cold atomic systems are now
allowing us to realize quasi-1D structures of
fermions as well, and will undoubtedly provide an excellent
experimental realization in which to study these problems.

\section{Conclusions and Perspectives}

1D and quasi-1D systems are a paradise for
QPTs. Due to the intrinsic critical nature
of interacting quantum systems, a pure 1D system can present a
set of instabilities, the simplest one being  associated
with the occurrence of one or several gaps in the systems. This occurs in the case of the Mott transition in 1D,
among many other examples.

Another important class of QPTs is driven
by the inter-chain coupling between 1D systems.
This is an especially important case given the direct
experimental relevance for several realizations, ranging from
naturally occurring materials to cold atoms in optical lattices. In
this case two main classes exist. If the 1D chain
is gapless, one usually finds a dimensional crossover between a
high temperature or high energy 1D regime and a
low temperature one dominated by the inter-chain coupling. For
spins or bosons, the low-temperature state is usually ordered.  Although this state is mostly an anisotropic
version of the 3D one, it can still retain some
special features coming from the quasi-1D
character. The case of fermions is more complex and the low-temperature phase is a higher-dimensional metal. If the 1D chains
are gapped one finds a deconfinement transition where the system
goes at zero temperature from a 1D gapped state to a high
dimensional ordered or gapless one, the latter occurring in
particular for fermions.  The nature of this transition in
fermionic systems is largely not understood and constitutes a
very challenging research field.

Although we have some of the tools and some
understanding of such transitions many crucial questions
remain. First, in contrast to the case of a purely 1D system for which we have a whole arsenal of
analytical and numerical tools to tackle the questions of such
transitions, the quasi-1D case is much more
difficult. Most of the numerical techniques are becoming very
inefficient, either due to their intrinsic limitations, such as the sign
problem for the fermions, or simply the large anisotropy of the
system that makes even well-controlled methods difficult to
apply. Clearly some new techniques are needed. On the
analytical side it is difficult to go beyond mean-field theory,
and thus to compute some of the correlation functions,
especially those involving directly the transverse degrees of
freedom. Going through the critical regime is also quite
challenging, even if we have a good idea of the various
phases. of special importance are the transverse transport,
the Hall or Nernst effect, and the propagation of some
transverse modes.

Some other topics are directly related to these issues and
present very challenging topics in themselves. The self-consistent dynamical mean-field approximation of  coupled LL
 replaces this problem by that of a LL in
equilibrium with an external bath.  The bath has drastic
consequences on the critical properties of the LL.  This type of
problem also occurs directly, either due to the presence of
external electrons or noise.  How to tackle such questions is
certainly one of the frontiers of our knowledge of 1D systems. Finally, all the examples of QCPs examined
in this chapter were based on a LL or a gapped phase as
the description of the 1D system. However, there are now
several identified 1D cases in which one has to go beyond the LL
paradigm to describe the physics of the systems. Understanding
the physics of such non-Luttinger liquids is a challenge in
itself, much as the understanding of non-Fermi liquids is. How
to go from the standard LL behavior to a non-LL one, or what
happens when non-LLs are coupled in a 3D lattice
is  a totally uncharted territory.

\textit{Acknowledgments} -- This work was supported in part by the Swiss NSF under MaNEP and Division II.

\bibliographystyle{prsty}
%\bibliography{totphys,qcp_1d}

\end{document}